\documentclass[aps,twocolumn,showpacs,superscriptaddress]{revtex4-1}
\usepackage{amsmath,amssymb}
\usepackage{graphics,graphicx}
\usepackage{dcolumn,bm}
\newcommand{\cu}
{\affiliation{Department of Physics, University of Calcutta, 
92 Acharya Prafulla Chandra Road, Kolkata 700009, India.}}
\newcommand{\snb}
{\affiliation{Department of Theoretical Sciences, S. N. Bose National Centre for Basic Sciences, 
JD Block, Sector-III, Salt Lake, Kolkata 700098, India.}}

\begin{document}

\title
{Agent based models for wealth distribution with preference in interaction}

\author{Sanchari Goswami}%
\email[Email: ]{sg.phys.caluniv@gmail.com}
\snb
\author{Parongama Sen}%
\email[Email: ]{psphy@caluniv.ac.in}
\cu


\begin{abstract}
We propose a set of conservative models in which agents exchange wealth with a preference in the choice of interacting 
agents in different ways. The common feature in all the models is that the temporary values of financial status 
of agents is a 
deciding factor for interaction. Other factors which may play important 
role are past interactions and wealth possessed by individuals. Wealth 
distribution, network properties and activity are the main quantities which have been studied. 
Evidence of phase transitions and other interesting features are presented. The results show that certain observations 
of real economic system can be reproduced by the models.  

\end{abstract}

\maketitle
\section{Introduction}
\label{intro}
One of the main objectives of several models in econophysics is to reproduce the Pareto tail or power-law tail 
in the wealth/income distribution in several economy \cite{Pareto:1897}. 
According to Pareto law, the probability that the income/wealth of an agent is equal to $m$ is given by,
\begin{equation}
 P(m) \sim m^{-(1+\nu)},
\end{equation}
where $\nu$ is called the Pareto exponent. The value of the exponent usually 
varies between $1$ and $3$~\cite{Mandelbrot:1960,EIWD, EWD05,ESTP,SCCC,Yakovenko:RMP,datapap}. 

Some of the models proposed to yield the above distribution are inspired by the kinetic theory of 
gases which derives the average macroscopic behaviour from the microscopic interactions between molecules. 
Agents can be regarded as molecules and 
a trading process can be regarded as an interaction between them.
In a typical trading 
a pair of traders exchange wealth, respecting local conservation of wealth in any trading
\cite{marjitIspolatov,Dragulescu:2000,Chakraborti:2000,Chatterjee:rev,Chakrabarti:2010,Chatterjee:2010}, 
similar to an elastic collision between molecules. Consequently, the total wealth remains conserved.
These agent based models have a microcanonical description and nobody ends up with negative wealth (i.e., debt is not allowed).
Thus, for two agents $i$ and $j$ with money $m_i(t)$ and $m_j(t)$ at time $t$, the general trading process is given by:
\begin{equation}
\label{mdelm}
m_i(t+1) = m_i(t) + \Delta m; \  m_j(t+1) = m_j(t) - \Delta m;
\end{equation}
time $t$ changes by one unit after each trading. The advantage of such models is that here dynamics at individual 
level can be studied. 
In a simple conservative model proposed by  Dragulescu and Yakovenko (DY model) \cite{Dragulescu:2000},  $N$ agents 
exchange wealth or money randomly keeping the total wealth $M$ constant. 
The steady-state ($t \rightarrow \infty$) wealth therefore follows a Boltzmann-Gibbs distribution:
$P(m)=(1/T)\exp(-m/T)$; $T=M/N$, 
a result which is robust and independent of the topology of the (undirected)
exchange space~\cite{Chatterjee:rev}.

An additional concept of \textit{saving propensity} was introduced first by Chakraborti 
and Chakrabarti~\cite{Chakraborti:2000} 
(CC model hereafter). Here, the agents save a fixed fraction $\lambda$ of their wealth
when  interacting  with another agent. 
This results in completely different types of wealth 
distribution curves, very close to Gamma distributions~\cite{Patriarca:2004,Repetowicz:2005,Lallouache:2010}
which fit well to empirical data for low and middle wealth regime~\cite{datapap}.
The model features are basically similar to Angle's work~\cite{Angle}.
In a later  model  proposed by Chatterjee et. al.~\cite{Chatterjee:2004} 
(CCM model hereafter) it was assumed that the saving propensity has a distribution, i.e., 
$\lambda$'s are now agent dependent
and this immediately led to a wealth distribution curve 
with a Pareto-like tail. 
Apart from these gas-like models, there are several other models of the wealth distribution. Some of these 
models depend on stochastic process \cite{Garl,Sornette} which cannot be realized as a real trading process. 
Another model is the Lotka-Volterra 
model where wealth of an agent at a particular step depends on their wealth in the previous step as well as the average 
wealth of all agents \cite{Solomon,Malcai}. The main problem in all these models is that here wealth exchange between agents is not allowed 
and therefore leads to a situation far from reality.

Although wealth distribution is one of the most important feature for 
which the models had been proposed, there are other interesting characteristics
of the market which a model should be able to reproduce. Financial institutions are seen to exhibit 
some interdependence and links are formed among them depending on several economic factors leading to network structure. 
In \cite{Allen, Babus} the problem of network
formation in a financial system have been addressed.
One can then study the network like features, e.g., the  kind of  clusters which 
are formed among agents and  the behaviour of the degree  distribution for better explanation of several economic 
phenomena. 
Some real data are available to this respect.
It has been shown that within a small interval of time 
most clusters are of size 2 \cite{tummi,tummi2} which can be termed as `dimerisation'. 
Another observation is regarding the  activity, i.e., the distribution of the volume of individual  trade 
that also  follows a power law with an exponent $\simeq 4.3$ \cite{Gabaix}.
These   features suggest that one needs to introduce some preference in the interaction between agents. 

In almost all the wealth exchange models, the interacting agents are selected randomly and any two agents have equal 
probability to interact.
In this paper, we incorporate preferential attachement to    agents for 
interaction as well as in the choice of  agents in some cases.
Such preferences need 
not be limited to geographically nearby neighbours. 
In \cite{manna1}, a preference in the selection of agents (according 
to their wealth) had been 
considered, however, the interacting agents were uncorrelated otherwise.

To obtain an optimized kinetic exchange model for trade, 
 several  features have to be incorporated.
Our basic assumption is that two agents will interact only when their 
wealths are ``close''. So in the simplest model, only such a feature is incorporated in an otherwise DY like model. 
More features have been added to obtain results closer to reality. 
In all 
the models wealth distribution, network features and other properties are studied. 

\section{Quantities Calculated}
\label{sec:2}
We consider kinetic exchange type models where the interactions are of DY type.
The simulation is done for a maximum of $N=1024$ agents. Initially the total money $M$ is distributed 
among the agents randomly. The stationary state is obtained after a typical 
relaxation time by checking the 
stability of the wealth distribution in the successive Monte Carlo (MC) steps, where one MC step 
is equivalent to $N$ pairwise interactions. The wealth distribution is obtained by averaging over a finite but large 
number of time steps. Finally the configurational averaging is done over a number of realizations 
to obtain the wealth distribution. 


Results for the following features have been presented in the paper:
\begin{enumerate}
 \item Wealth Distribution: $P(m)$ (already introduced in sec \ref{intro}),
 \item Degree Distribution:
The number of agents with whom one particular agent interacts within one MC time step, averaged 
over all time step is the degree of an 
agent. $D(k)$ denotes the probability that an agent has degree $k$.
\item Activity Distribution:
Activity distribution is defined as the number of transactions made by one individual in one MC
timestep, averaged over all timesteps. We use $Q(A)$ to denote the 
activity distribution.
\item Average degree with wealth $m$:
$d(m)$, the average degree of an agent with money $m$ is also calculated to investigate 
whether the degree is correlated to wealth.
\end{enumerate}

In all the cases, we have taken  $ M = \sum_{i=1}^{N} m_i $ to be equal to   $N$.

\section{Models and Results}
\label{sec:3}
\subsection{Model A}
In model A, the only criteria that an interaction between two agents will take place is that they should be 
financially close. The 
probability of interaction between agents $i$ and $j$ is taken as
\begin{equation}
 P_{ij} \propto |m_i-m_j|^{-\alpha}.~~~~~\alpha >0.
\end{equation}
Note that, it may happen that wealth of two agents $i$ and $j$ are equal, i.e., $m_i=m_j$. 
In that case, it was considered that interaction between $i$ and $j$ would occur as a sure event. However as 
$m$ is continuously varying such cases are extremely rare.

\begin{figure}
\includegraphics[width=7cm, angle=-90]{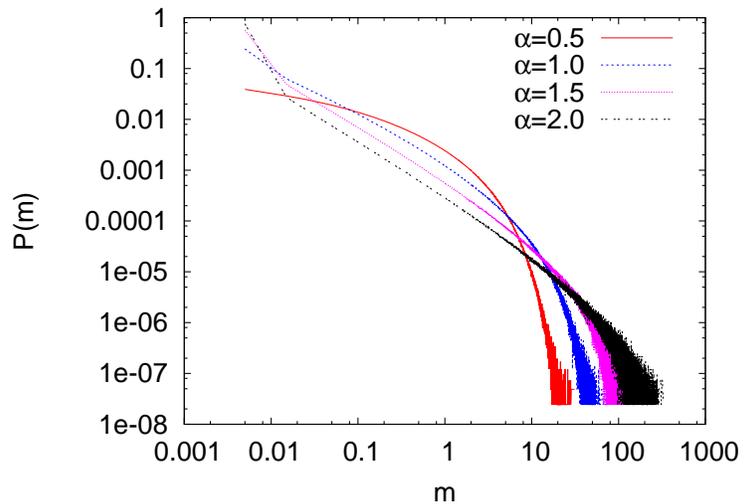}
\caption{(Color online) Plot of  $P(m)$ as a function of $m$ for model A with $N = 1024$. Total money $M=N$.
}
 \label{mnydist_A}
\end{figure}

\begin{figure}
\includegraphics[width=6cm, angle=-90]{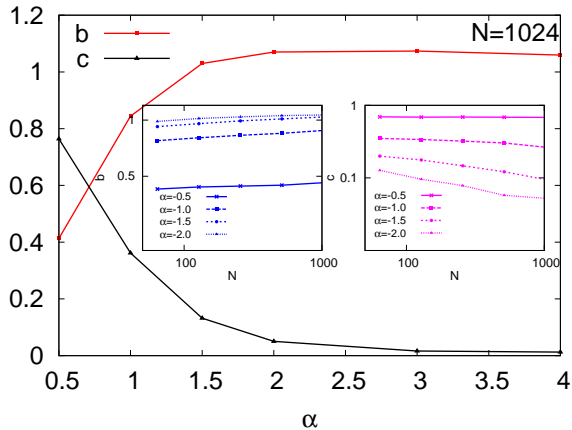}
\caption{(Color online) Variation of  $b$ and $c$ with $\alpha$ for model A for N=1024. Inset shows
(left) variation of  $b$ with $N$ for model A and (right) variation of  $b$ with $N$ for model A.
}
 \label{bc_alpha}
\end{figure}

Wealth distribution for model A for extreme values of $\alpha$ are as follows :
\begin{enumerate}
 \item For $\alpha=0$, we get back the DY model.
 \item When $\alpha >> 1$, the tail of the wealth distribution has a power law form.
\end{enumerate}
The wealth distribution for different values of $\alpha$ are shown in Fig. \ref{mnydist_A}.
It is seen from the figure that the plots have the general form $am^{-b}\exp(-cm)$. The variations of $b$ and $c$ with 
$\alpha$ are shown in Fig. \ref{bc_alpha} and those with $N$ are shown in the inset.
Variation of $c$ with  $N$ indicates that it vanishes at large values of $\alpha$ in the 
thermodynamic limit. The value of $b$ 
increases with increasing $\alpha$ and $N$. For $\alpha \gtrsim 2$ it is close to $1$. 
One can safely conclude that for $\alpha > 2$,
the exponential cut-off vanishes. However, the corresponding Pareto exponent is rather small ($\nu = b-1 \sim 0.1$).

\begin{figure}
\includegraphics[width=6cm, angle=-90]{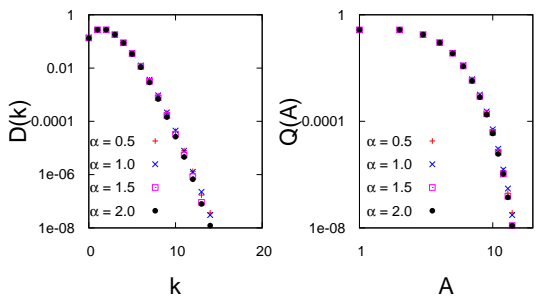}
\caption{(Color online) Plot of D(k) and Q(A) for model A with $N = 1024$. 
}
 \label{deg_act}
\end{figure}

Degree distribution $D(k)$ for model A has an 
exponential form and does not change appreciably with $\alpha$ as is shown in the left panel of Fig. \ref{deg_act}. 


In the right panel of Fig. \ref{deg_act} the activity distribution $Q(A)$ for model A is shown. 
It has an exponential form that does not change with $\alpha$. 

\begin{figure}
\includegraphics[width=6cm, angle=-90]{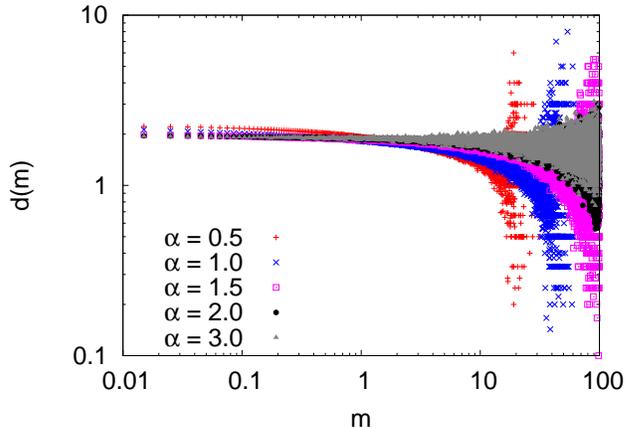}
\caption{(Color online) Plot of  $d(m)$ as a function of $m$ for model A. 
}
 \label{avdeg_mny}
\end{figure}

Average degree of an agent with wealth $m$ is represented by $d(m)$ and is 
shown in Fig. \ref{avdeg_mny}. 
It is seen that $d(m)$ is independent of the wealth possessed by an individual; more so
for larger values of $\alpha$. However, for large values of $m$ ($m > 10$), there is appreciable fluctuation.

\subsection{Model B}
In model B, in addition to the assumption that transactions are more probable for agents who are financially close to each 
other, it is assumed that probability of transaction increases with past number of interactions. Probability of interaction 
between $i$ and $j$ taken in model B is,
\begin{equation}
 P_{ij} \propto |m_i - m_j|^{-\alpha} (c_{ij}+1)^{\gamma},
\end{equation}
where $c_{ij}$ is the number of interactions which have taken place already 
between $i$ and $j$. The factor $1$ is added to $c_{ij}$ to ensure that 
two persons who have not traded with each other yet can still interact. 

\begin{figure}
\includegraphics[width=6cm, angle=-90]{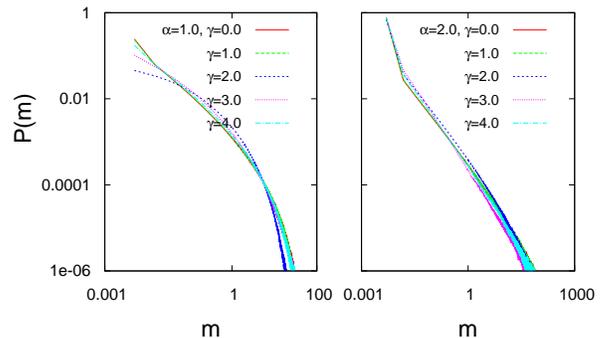}
\caption{(Color online) Plot of  $P(m)$ as a function of $m$ for model B with parameters $\alpha = 1.0$ and $2.0$, 
$\gamma = 0.0, 1.0, 2.0, 3.0$ and $N = 1024$. 
}
 \label{mnydist_B}
\end{figure}

 The wealth distributions for two different values of $\alpha$ for various values of 
 $\gamma$ are shown in Fig. \ref{mnydist_B}.
For $\alpha \geq 2$ one gets the power law tail but the 
corresponding values of $\nu$ are still quite  small ($\mathcal {O}(0.1))$. 
Some of the values of $\nu$ obtained  for different chosen 
sets of parameters for model B are shown in Table \ref{table}.


\begin{figure}
 \includegraphics[width=6cm, angle=-90]{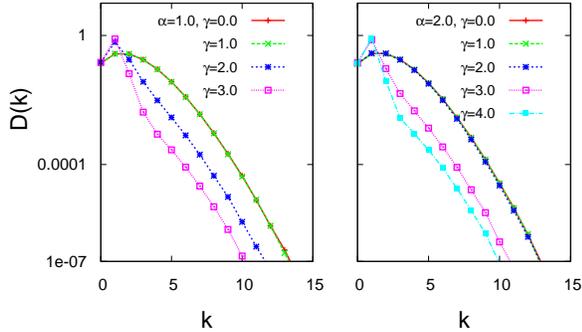}
\caption{(Color online) Plot of  degree distribution $D(k)$ as a function of degree $k$ for model B with parameters 
$\alpha = 1.0$ and $2.0$, $\gamma = 0.0, 1.0, 2.0, 3.0$ and $N = 1024$. 
}
 \label{DD_B}

\end{figure}
\begin{figure}
 \includegraphics[width=6cm, angle=-90]{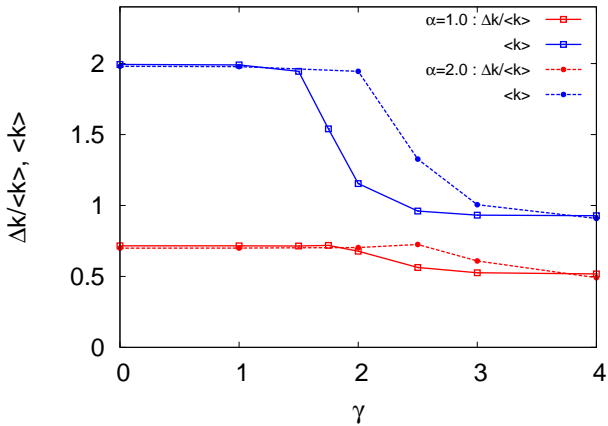}
\caption{(Color online) Plot of  mean degree $\langle k \rangle$ and fluctuation $\frac{\Delta k}{\langle k \rangle}$. 
}
 \label{fluc}
\end{figure}
\begin{figure}
 \includegraphics[width=6cm, angle=0]{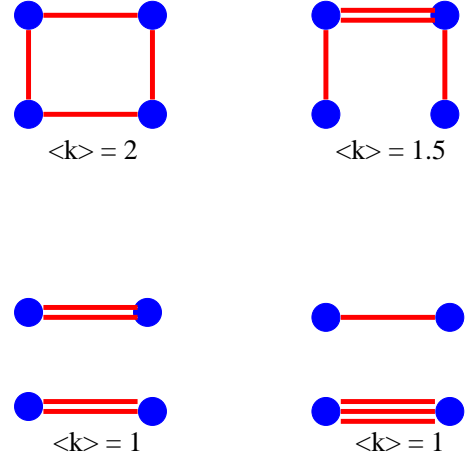}
\caption{(Color online) Example of decrease of average degree with four nodes and four links. It is shown 
how average degree decreases as repeated interaction between agents take place leading to dimerisation. 
}
 \label{dimer}
\end{figure}

The degree distribution is shown in Fig. \ref{DD_B}. The mean degree decreases for higher values of $\gamma$. 
The mean degree $\langle k \rangle$ and fluctuation $\frac{\Delta k}{\langle k \rangle}$ 
are plotted against $\gamma$ (Fig. \ref{fluc}). 
It shows an interesting feature:  $\langle k\rangle$ has a value equal to $2$ for small $\gamma$ and 
equal to $1$ for larger values of $\gamma$. Variation of $\langle k\rangle$ from $2$ to $1$ is obtained 
over a narrow region of $\gamma$ values. The decrease of mean degree can be interpreted in the following way: 
as $\gamma$ increases, interaction involving the same pair of agents is repeated and
effectively a dimerisation takes place.
Similar dimers and small clusters have been observed by Tumminello et. al. \cite{tummi2} for agents
in stock market data.
Crossover to a dimerised state occurs as $\gamma$ is increased.
A simple example of how dimerisation affects the average degree is shown in Fig. \ref{dimer}.

Activity distribution is similar to model A and does not show any special feature.

\begin{figure}
\includegraphics[width=6cm, angle=-90]{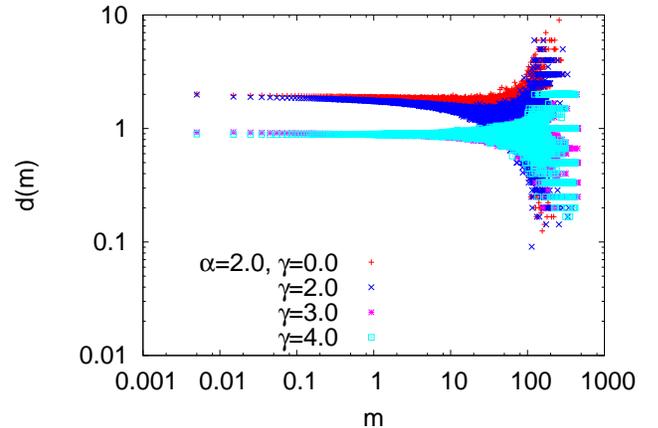}
\caption{(Color online) Plot of  $P_d(m)$ as a function of $m$ for model B with parameters $\alpha = 2.0$,
$\gamma = 0.0, 2.0, 3.0, 4.0$  and $N = 1024$. 
}
 \label{avdeg_mny_B}
\end{figure}

The data for $d(m)$, average degree of an agent with wealth $m$ is 
shown in Fig. \ref{avdeg_mny_B}. 
It is seen that $d(m)$ is again independent of the amount of wealth possessed by an agent (for $m \lesssim 1$) as in model A 
but can assume only two different values close to  1 and 2.  $d(m) \simeq 1$ 
 corresponds  to   a 
 larger $\gamma$ value when dimerisation occurs. 

\begin{table}[t]
\centering
\begin{tabular}{|l|c|c|c|c|}
\hline
Type of the model & $\alpha$ & $\beta$ & $\gamma$ & $\nu$ \\
\hline \hline
Model B & $2.0$ & $0.0$ & $1.0$ & $0.088$ \\
\cline{2-5}
 & $2.0$ & $0.0$ & $2.0$ & $0.096$ \\
\cline{2-5}
 & $2.0$ & $0.0$ & $3.0$ & $0.279$ \\
\cline{2-5}
 & $2.0$ & $0.0$ & $4.0$ & $0.174$ \\
\hline \hline
Model C & $3.0$ & $1.0$ & $0.0$ & $0.798$ \\
\cline{2-5}
& $3.0$ & $2.0$ & $0.0$ & $1.432$ \\
\cline{2-5}
& $3.0$ & $3.0$ & $0.0$ & $2.134$ \\
\hline \hline
Model D & $2.0$ & $1.0$ & $1.0$ & $0.671$\\
\cline{2-5}
& $2.0$ & $1.0$ & $2.0$ & $0.400$\\
\cline{2-5}
& $2.0$ & $1.0$ & $3.0$ & $0.091$\\
\cline{2-5}
& $3.0$ & $1.0$ & $1.0$ & $0.792$\\
\cline{2-5}
& $3.0$ & $1.0$ & $2.0$ & $0.519$\\
\cline{2-5}
& $3.0$ & $1.0$ & $3.0$ & $0.196$\\ 
\cline{2-5}
& $3.0$ & $3.0$ & $2.0$ & $2.34$\\
\hline
\end{tabular}
\caption{Different values of Pareto Exponent for different combinations of $\alpha$, $\beta$ and $\gamma$.}
\label{table}
\end{table}

\subsection{Model C}
In model C, the first agent $i$ is chosen with a probability $p_i = m_{i}^{\beta}$, where $\beta$ is a 
parameter. Chakraborty et. al. \cite{manna1} in a recent paper used such a preferential 
selection rule using a pair of continuously tunable 
parameters upon traders with distributed saving
propensities and were able to reproduce the trend of  enhanced rates
of trading of the rich. The wealth distribution was found to follow Pareto law.
However, in model C, we choose only the first agent with a preferential selection rule. The second agent is chosen  
with higher probability when she is financially close to the first as in models A and B. 
The interaction in model C occurs 
with a probability $P_{ij}$, given by,
\begin{equation}
 P_{ij} \propto m_{i}^{\beta} |m_i - m_j|^{-\alpha}.
\end{equation}
In effect, both the interacting agents 
are rich for higher value of $\alpha$. 

\begin{figure}
 \includegraphics[width=6cm, angle=-90]{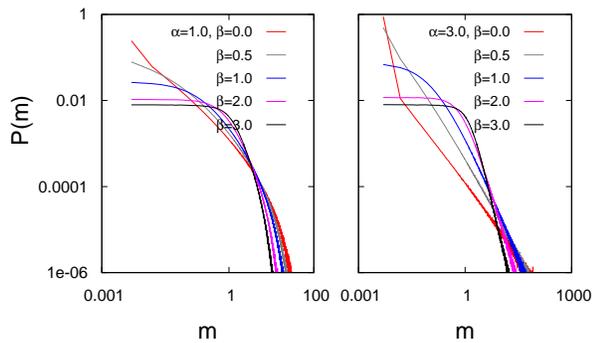}
\caption{(Color online) Plot of  $P(m)$ as a function of $m$ for model C with $N = 1024$ and 
parameters $\alpha = 1.0,3.0$ and for 
$\beta = 0.0, 0.5, 1.0, 2.0, 3.0$. 
}
 \label{mnydist_C}
\end{figure}

The wealth distributions for two different values of $\alpha$ and various values of $\beta$ are shown in 
Fig. \ref{mnydist_C}.

It is seen that the wealth distribution is sufficiently altered for $\beta \neq 0$; a plateau/flat region is found for small $m$,
and a power law region for a narrow range of $m$ follows it. It can be interpreted in the following way: as 
selection of the agents depend on their wealth,
many agents may not interact at all. Now, poorer agents have less probabilities to interact. 
Thus the wealth distribution is almost flat up to a certain value of 
$m$. As agents become richer, they interact more and the form of wealth distribution   
shows variation with $m$. The exponent $\nu$ for the power law region is quite important 
here, because now it has an appreciable value $\nu \gtrsim 1$. 
As $\beta$ increases the value of $\nu$ also increases. For example, for a chosen set 
of parameters 
$\alpha = 2.0, \beta = 3.0$, the exponent $\nu$ has a value close to $2$. Pareto exponents for model C are 
shown in table \ref{table}.

\begin{figure}
 \includegraphics[width=6cm, angle=-90]{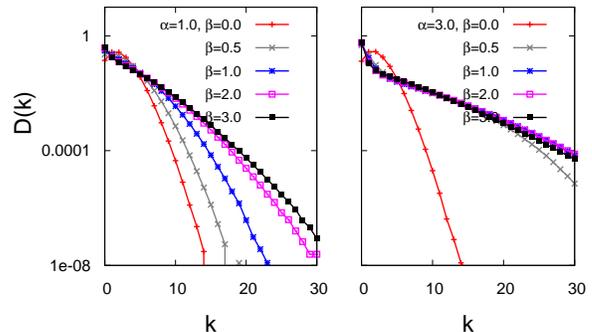}
\caption{(Color online) Plot of  degree distribution $D(k)$ as a function of degree $k$ for model C with $N = 1024$ and 
parameters $\alpha = 1.0$ and $3.0$ and for 
$\beta = 0.0, 0.5, 1.0, 2.0, 3.0$. 
}
 \label{DD_C}

\end{figure}

Degree distribution in model C is more spread out compared to models A and B, as shown in Fig. \ref{DD_C}. The plot 
indicates that for large $\alpha$, whatever value of $\beta$ we choose (except zero), we have a large number 
of agents with high degrees. 

\begin{figure}
 \includegraphics[width=6cm, angle=-90]{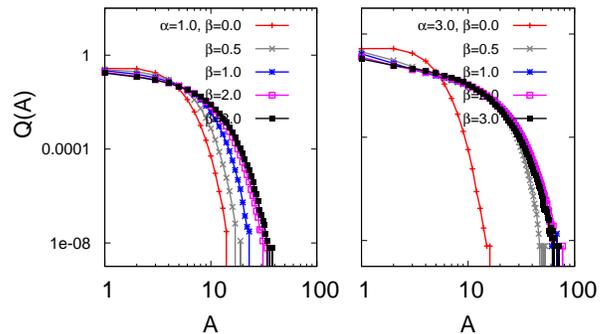}
\caption{(Color online) Plot of  activity distribution $Q(A)$ for model C with $N = 1024$ and 
parameters $\alpha = 1.0$ and $3.0$ and for 
$\beta = 0.0, 0.5, 1.0, 2.0, 3.0$. 
}
 \label{act_C}
\end{figure}

Here the activity distribution (Fig. \ref{act_C}) shows a distinct parameter dependence unlike 
models A and B. For large $\alpha$ and nonzero value of $\beta$, there is a considerably higher probability of large activity. 
As $\alpha$ and $\beta$ simultaneously help the interaction between rich agents to occur, for large value of $\alpha$ and nonzero $\beta$, 
interaction is 
limited within a \textquoteleft rich\textquoteright group. These agents 
therefore enjoy large activity which is the reason why $Q(A)$ 
is nonzero for much larger values of $A$.

\begin{figure}
\includegraphics[width=6cm, angle=-90]{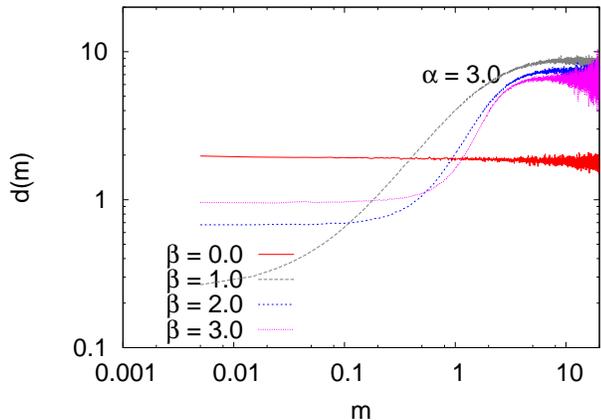}
\caption{(Color online) Plot of  $d(m)$ as a function of $m$ for model C with $N = 1024$ and 
parameters $\alpha = 3.0$, 
$\beta = 0.0, 0.5, 1.0, 2.0, 3.0$. 
}
 \label{avdeg_mny_C2}
\end{figure}

Average degree of an agent with wealth $m$, i.e., $d(m)$ is 
shown in Fig.  \ref{avdeg_mny_C2}. 
It shows a different behaviour compared to
models A and B. It is no longer a flat distribution. Richer agents have more
neighbours as they have a priority in interactions resulting in an increasing trend in  $d(m)$ with $m$ for large $m$.

\section{Model D}
In model D, we consider all the features contributed by the parameters $\alpha, \beta$ and $\gamma$.
Here $P_{ij}$ can be written as,
\begin{equation}
 P_{ij} \propto  |m_i - m_j|^{-\alpha} m_{i}^{\beta} (c_{ij}+1)^{\gamma}
\end{equation}
where $\beta = 0, \gamma = 0$ gives model A; $\beta = 0$ gives model B and $\gamma = 0$ gives model C.

\begin{figure}
\includegraphics[width=6cm, angle=-90]{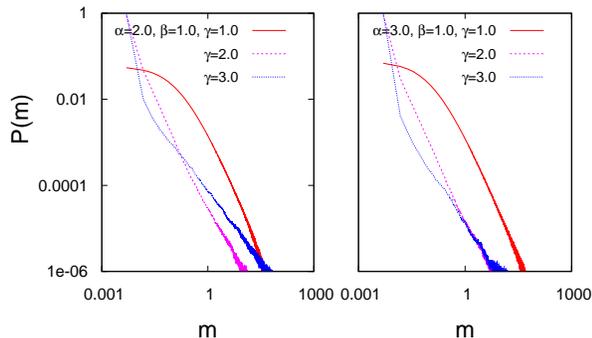}
\caption{(Color online) Plot of  $P(m)$ as a function of $m$ for model D with $N = 1024$ and 
parameters $\alpha = 2.0, 3.0, \beta = 1.0$ and $\gamma = 1.0, 2.0, 3.0$. 
}
 \label{mny_D}
\end{figure}

The corresponding wealth distribution for model D is shown in Fig. \ref{mny_D}. Note that as we increase 
$\gamma$ beyond $1$, the flat region disappears. 
With the presence of all three parameters, the value of the exponent $\nu$ is close to $1$ when $\gamma$ is small 
and decreases as $\gamma$ increases. The different values of $\nu$ for different combination 
of the parameter values are shown in Table \ref{table}.

\begin{figure}
\includegraphics[width=6cm, angle=-90]{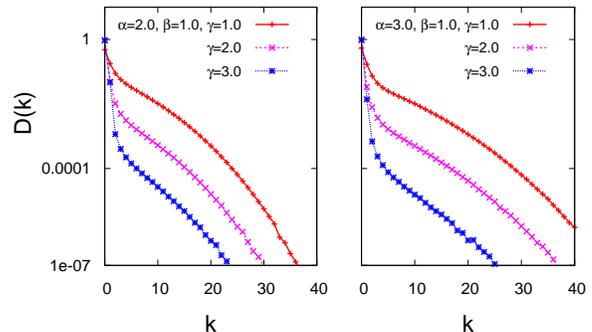}
\caption{(Color online) Plot of  $D(k)$ as a function of $k$ for model D with $N = 1024$ and 
parameters $\alpha = 2.0, 3.0$, $\beta = 1.0$ and for 
$\gamma = 1.0, 2.0, 3.0$. 
}
 \label{DD_D}
\end{figure}
\begin{figure}
\includegraphics[width=6cm, angle=-90]{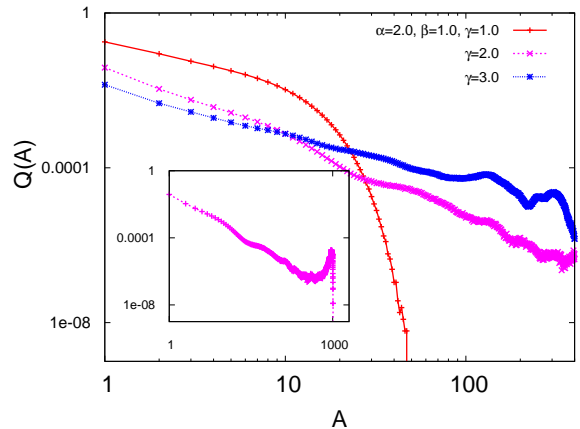}
\caption{(Color online) Plot of  activity distribution $Q(A)$ for model 
D with $N = 1024$ and 
parameters $\alpha = 2.0$, $\beta = 1.0$ and for 
$\gamma = 1.0, 2.0, 3.0$. The same plot for $\alpha=2.0, \beta=1.0, \gamma=2.0$ is shown for 
the whole range of activity in the inset showing the condensate like behaviour at large values of $A$.
}
 \label{act_D}
\end{figure}
\begin{figure}
\includegraphics[width=6cm, angle=-90]{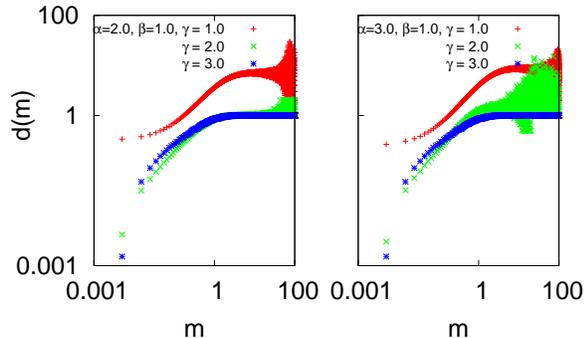}
\caption{(Color online) Plot of  $d(m)$ as a function of $m$ for model D with $N = 1024$ and 
parameters $\alpha = 1.0$, $\beta = 3.0$ and for 
$\gamma = 1.0, 2.0, 3.0$. 
}
 \label{avdeg_mny_D}
\end{figure}
Here degree distribution has maximum  value for $k=0$ and then drops off suddenly to a low value as 
shown in Fig. \ref{DD_D}. The fall is sharper as $\gamma$ 
increases. Also note that with increasing $\alpha$, the degree distribution is more spread out as is also seen in 
Fig. \ref{DD_C}. A feature similar to dimerisation as in model B is also observed here, but the average degree
varies from $2$ to $0$ over a region of $\gamma$ unlike model B where the variation is from $2$ to $1$.

A striking feature is observed for activity distribution of model D. It can be seen that 
here with all the three parameters present, unless $\gamma$ is very small, 
the activity distribution shows power-law behaviour 
as shown in Fig. \ref{act_D}. The corresponding exponent is dependent on the value of the parameters and 
in general around 3 which is somewhat less than the observed 
value \cite{Gabaix}. 
 The power law behaviour of $Q(A)$ signifies the presence of a few agents with large amount 
of activity - evidently the rich agents have these property. In fact, for higher values of $\alpha$ and $\beta$, 
this effect is enhanced leading to the existence of a local peak at $A >> 1$. However, the height of this peak is 
much lesser compared to $Q(A=1)$. 
This ``condensation'' type behaviour becomes more prominent for larger values of $\alpha$.
Average degree $d(m)$ of an agent with wealth $m$ shows features similar to model C and is shown in Fig. \ref{avdeg_mny_D}.

\section{Comparison with real data}
While modelling a particular system, e.g., as in \cite{GIori}, one may calibrate the numerical
simulations of the model with real data. Even for a general model, it is important that 
the exponent values of the relevant quantities obtained are comparable to 
real data. We have extracted the Pareto exponent and the exponent for the activity 
distribution wherever possible for the models proposed in this paper.

\begin{figure}
\includegraphics[width=8cm, angle=0]{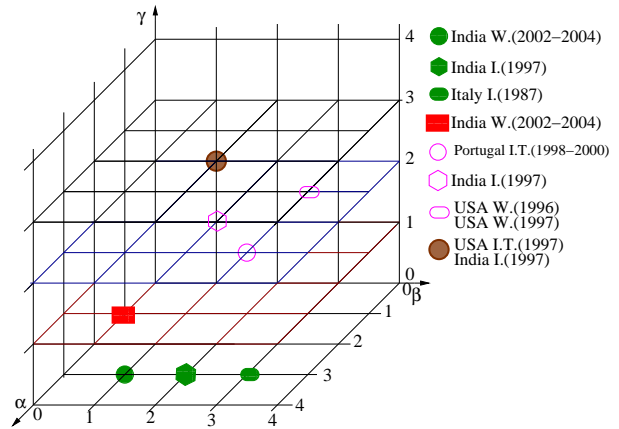}
\caption{(Color online)3-\textit{d} plot for the suitable values of $\alpha$, 
$\beta$ and $\gamma$ yeilding $\nu$ values corresponding to different coutries. 
}
 \label{3d}
\end{figure}

While considering real data, it is possible only to compare the Pareto exponents obtained from the different models. 
We therefore check for the values of $\alpha$, 
$\beta$ and $\gamma$ which give us a Pareto exponent $\nu$ comparable to the real data of several countries which
appear in \cite{Pete_Rich}. In Fig. \ref{3d}, we show in a 3-\textit{d} plot the suitable values of $\alpha$, 
$\beta$ and $\gamma$ yeilding $\nu$ values corresponding to different coutries. However, there is a word of caution - 
a particular real Pareto value may be obtained by more than one combination of $\alpha$, 
$\beta$ and $\gamma$ and one should not try to interpret the values shown in Fig. \ref{3d} to be optimum.
\section{Summary and Discussions}
To summarize, we have studied different wealth exchange models where agents interact via DY type interaction in 
addition to the fact the interactions among the agents 
are now preferential. For all the models we assume that two agents will interact only when they are ``closely'' located 
in the wealth space. This is controlled by the parameter $\alpha$ which is taken to be non-zero in all the models. The introduction of $\alpha$ leads to 
a power law behaviour in $P(m)$ above a certain value of $\alpha$ even without 
considering other factors like saving.
To mimic the real situation we have also incorporated other parameters $\beta$ and $\gamma$. $\gamma$ takes care 
of the ``memory'' that a pair of agents have interacted already; probability of interaction 
increases with the number of past interactions  
controlled by the parameter $\gamma$.
The parameter $\beta$ helps to select the agents with a probability proportional to their wealth.
Although our prime concern is the wealth distribution, the issue of network formation has also 
been addressed by considering some fundamental network properties. In several earlier works, the question of network
 formation
in financial systems has been considered \cite{Allen, Babus}.

With only $\alpha \neq 0$, one can get a power law decay in $P(m)$
with a Pareto exponent $\nu \approx 0.1$. With the introduction of either $\beta$ and/or $\gamma$, one still gets the 
power law decay but the value of  $\nu$  shows  drastic change. In principle it is possible to obtain a specific 
value of $\nu$ by properly tuning the parameters. 
When  $\gamma$ is nonzero, an additional feature of dimerisation, observed 
in real data, appears in the results. However, the average degree does not show any significant dependence on the wealth 
possessed by an agent in models A and B (i.e. $\beta =0$). When $\beta$ is  nonzero, 
the new feature which is observed is 
the nontrivial dependence of the average degree on the wealth of an agent, there being a distinct 
nonlinear increasing trend for higher values of $m$. 

When all the three parameters are present (model D), one gets a power law for $\alpha \geq 2$ as
in all the other models (A, B and C), and once again it is possible to generate various $\nu$ values
by different combinations of $\alpha, \beta$ and $\gamma$ (Table \ref{table}). 
For model D, 
another  desirable feature is  obtained in addition to dimerisation and nontrivial variation of $d(m)$ with $m$. 
This is the power law observed in the activity distribution. 

One has to carefully choose the values of $\alpha$, $\beta$ and $\gamma$ so as to achieve optimum behaviour in model D.  
We find that for $\alpha \sim 2, \beta \sim 1$ and $\gamma \sim 2$, the features become closest to reality. For example, 
making $\gamma$ large and $\beta$ small, the value of $\nu$ decreases, while for smaller values of $\gamma$ 
the activity $Q(A)$ does not show a power law behaviour. If $\beta$ is chosen as $>1$, the plateau region in $P(m)$ extends 
over a larger region of $m$ which is an undesirable feature. While $\alpha >1$ ensures a power law behaviour in $P(m)$, 
making $\alpha$ too large in model D leads to enhanced condensation behaviour.  
It should be mentioned that finite size effects 
for all cases are negligible for system size for which the results are reported.

However the  problem with these optimum values of the is that the Pareto exponent $\nu$ in this case is rather small ($\sim 0.4$). The activity distribution also has an exponent 
smaller than the observed one. To obtain  better values of these exponents 
 one might try further fine tunings and incorporate features like saving.

\medskip

Acknowledgement: Discussion with S. S. Manna during the initial formulation of the problem is acknowledged. PS is thankful to CSIR grant for financial support.


\end{document}